\documentclass{procs}

\usepackage{amsmath}
\usepackage{enumitem}
\usepackage{cite}

\def\be{\begin{equation}}
\def\ee{\end{equation}}
\def\bea{\begin{eqnarray}}
\def\eea{\end{eqnarray}}

\newcommand{\Photo}{}

\newcommand{\gev}{\,\, \mathrm{GeV}}
\newcommand\MHp{M_{H^\pm}}
\newcommand{\br}{\text{BR}}
\newcommand{\mnSSM}{\ensuremath{\mu\nu\mathrm{SSM}}}
\newcommand{\tb}{\tan\beta}

\begin{document}
\begin{flushright}
\mbox{}
IFT--UAM/CSIC--20-041~\footnote{Talk presented at the IAS Program on High Energy Physics (HEP) 2020.}  \\
DESY 20--047
%arXiv:20mm.nnnnn [hep-ph]
\end{flushright}
\vspace*{1cm}

\title{The ``96 GeV excess'' at the LHC}

\author{T. Biek\"otter}
\addresss{DESY, Notkestrasse 85,
D-22607 Hamburg, Germany}
%\address{Instituto de F\'isica Te\'orica (UAM/CSIC) \& \\
%Departamento de F\'isica Te\'orica, Universidad Aut\'onoma de Madrid\\}

\author{M. Chakraborti}
\addresss{IFT (UAM/CSIC), Universidad Aut\'onoma de Madrid, 
Cantoblanco, E-28048, Spain}

\author{S. Heinemeyer}
\address{IFT (UAM/CSIC), Universidad Aut\'onoma de Madrid
Cantoblanco, E-28048, Spain\\
Campus of International Excellence UAM+CSIC, Cantoblanco, E-28049,
Madrid, Spain\\
Instituto de F\'isica de Cantabria (CSIC-UC), E-39005 Santander, Spain}

\maketitle\abstracts{
The CMS collaboration reported an intriguing $\sim 3 \, \sigma$ (local) excess
at $96\;$GeV in the light Higgs-boson search in the diphoton decay mode.
This mass coincides with a $\sim 2 \, \sigma$ (local) excess in the
$b\bar b$ final state at LEP.
We briefly review the proposed combined interpretations for the two excesses.
In more detail we review the interpretation of this
possible signal as the lightest Higgs boson in the 2 Higgs Doublet Model
with an additional real Higgs singlet (N2HDM). We show which channels
have the best prospects for the discovery of additional Higgs bosons at
the upcoming Run~3 of the LHC. 
}

\section{Introduction}

The Higgs boson discovered in 2012 by ATLAS and
CMS~\cite{Aad:2012tfa,Chatrchyan:2012xdj} is so far consistent with
the existence of a Standard-Model~(SM) Higgs
boson~\cite{Khachatryan:2016vau} with a mass of $\sim
125\,$GeV. However, the experimental uncertainties on the Higgs-boson
couplings are (if measured already) at the precision of $\sim 20\%$,
so that there is room for Beyond Standard-Model (BSM)
interpretations. Many theoretically well motivated extensions of the
SM contain additional Higgs bosons. In particular, the presence of
Higgs bosons lighter than $125\,$GeV is still possible. 

Searches for light Higgs bosons have been performed at LEP, the
Tevatron and the LHC. Besides the SM-like Higgs boson at $125\,$GeV no
further detections of scalar particles have been reported. However,
two excesses have been seen at LEP and the LHC at roughly the same
mass, hinting to a common origin of both excesses via a new particle
state. LEP observed a $2.3\, \sigma$ local excess 
in the~$e^+e^-\to Z(H\to b\bar{b})$
searches\,\cite{Barate:2003sz}, consistent with a
scalar of mass ~$\sim 98\,$GeV, where the mass
resolution is rather imprecise due to the
hadronic final state. The signal strength
was extracted to be $\mu_{\rm LEP}= 0.117 \pm 0.057$.
The signal strength $\mu_{\rm LEP}$ is the
measured cross section normalized to the SM expectation
assuming a SM Higgs-boson mass at the same mass.

CMS searched for light Higgs bosons in the diphoton
final state. Run II\,\cite{Sirunyan:2018aui} results
show a local excess of $\sim 3\, \sigma$ at
$\sim 96\,$GeV, and a similar excess of $2\, \sigma$
at roughly the same mass~\cite{CMS:2015ocq} in Run~I.
Assuming dominant gluon fusion production the
excess corresponds to $\mu_{\rm CMS}=0.6 \pm 0.2$.
First Run\,II~results from~ATLAS
with~$80$\,fb$^{-1}$ in the diphoton final state turned
out to be weaker than the corresponding CMS results, see, e.g., Fig.~1
in~\cite{Heinemeyer:2018wzl}.

Here we first briefly review the models that have been proposed to
explain the two excesses together. Then we concentrate on the
2~Higgs Doublet Model with an additional real Higgs singlet (N2HDM).
It is shown that the type~II
and type~IV (flipped) of the N2HDM can perfectly accommodate both
excesses simultaneously, while being in agreement with all
experimental and theoretical constraints. The excesses are most
easily accommodated in the type II~N2HDM, which resembles the
Yukawa structure of supersymmetric models. We show which channels
have the best prospects for the discovery of additional Higgs bosons at
the upcoming Run~3 of the LHC. This complements our original N2HDM
analysis~\cite{Biekotter:2019kde}, together with previous proceedings
extending the original work with additional ILC/LHC/EWPO
analyses~\cite{Biekotter:2019drs}, as well as $e^+e^-$/ILC specific
analyses~\cite{Biekotter:2020ahz}. 

%%%%%%%%%%%%%%%%%%%%%%%%%%%%%%%%%%%%%%%%%%%%%%%%%%%%%%%%%%%%%%%%%%%%%%%%%%%%%%
%%%%%%%%%%%%%%%%%%%%%%%%%%%%%%%%%%%%%%%%%%%%%%%%%%%%%%%%%%%%%%%%%%%%%%%%%%%%%%

\section{The experimental data and possible BSM interpretations}
\label{sec:models}

LEP reported a $2.3\,\sigma$ local excess
in the~$e^+e^-\to Z(H\to b\bar{b})$
searches\,\cite{Barate:2003sz}, which would be consistent with a
scalar mass of~$\sim 98 \gev$ (but due to the final state the 
mass resolution is rather coarse). The ``excess'' corresponds to 
\begin{equation}
\mu_{\rm LEP}=\frac{\sigma\left( e^+e^- \to Z \phi \to Zb\bar{b} \right)}
			   {\sigma^{SM}\left( e^+e^- \to Z H_{\rm SM}
			   		\to Zb\bar{b} \right)}
			  = 0.117 \pm 0.057 \; ,
\label{muLEP}
\end{equation}
where the signal strength $\mu_{\rm LEP}$ is the
measured cross section normalized to the SM expectation,
with the SM Higgs-boson mass at $\sim 98\gev$.
The value for $\mu_{\rm LEP}$ was extracted in \cite{Cao:2016uwt}
using methods described in \cite{Azatov:2012bz}.

CMS~Run\,II searches for Higgs bosons decaying in the diphoton
final state show a local excess of~$\sim 3\,\sigma$ around
$\sim 96 \gev$~\cite{CMS:2017yta}, with a similar excess
of~$2\,\sigma$ in the Run\,I data at a comparable mass.  
In this case the ``excess'' corresponds to (combining 7, 8 and 13~TeV data)
\begin{equation}
\mu_{\rm CMS}=\frac{\sigma\left( gg \to \phi \to \gamma\gamma \right)}
         {\sigma^{\rm SM}\left( gg \to H_{\rm SM} \to \gamma\gamma \right)}
     = 0.6 \pm 0.2 \; .
\label{muCMS}
\end{equation}
First Run\,II~results from~ATLAS
with~$80$\,fb$^{-1}$ in the~$\gamma\gamma$~searches below~$125$\,GeV were
published~\cite{ATLAS:2018xad}. While no significant excess above
the~SM~expectation was observed in the mass range between $65$~and
$110 \gev$, it was found that 
the limit on cross section times branching ratio obtained in the
diphoton final state by ATLAS is not only well above $\mu_{\rm CMS}$,
but even weaker than the corresponding upper limit obtained by CMS at 
$\sim 96 \gev$~\cite{Heinemeyer:2018wzl} (see Fig.~1 therein).

\bigskip
Several analyses attempted to explain the {\em combined} ``excess'' of
LEP and CMS in a variety of BSM models%
\footnote{More analyses attempted to explain one of the two
  ``excesses'', but we will not discuss these further here.}%
. To our knowledge explanations exist in the following frameworks
(see also \cite{Heinemeyer:2018wzl,Heinemeyer:2018jcd,Richard:2020jfd}):
\begin{itemize}
\item Higgs singlet with additional vector-like
  matter, as well as Type-I 2HDM~\cite{Fox:2017uwr}.
\item Radion model~\cite{Richard:2017kot}.
\item Type-I 2HDM with a moderately-to-strongly
  fermiophobic CP-even Higgs~\cite{Haisch:2017gql}.
\item \mnSSM\ with one~\cite{Biekotter:2017xmf} and three
  generations~\cite{Biekotter:2019gtq} of right-handed neutrinos.
\item Higgs associated with the breakdown of an $U(1)_{L_\mu L_\tau}$
  symmetry~\cite{Liu:2018xsw}. 
\item Various realizations of the
  NMSSM~\cite{Domingo:2018uim,Choi:2019jts}. 
\item Higgs inflation inspired $\mu$NMSSM~\cite{Hollik:2018yek}.
\item NMSSM with a seesaw extension~\cite{Cao:2019ofo}.
\item N2HDM~\cite{Biekotter:2019kde}, as will be discussed below.
\item Minimal dilaton model~\cite{LiuLiJia:2019kye}.
\item
SM extended by a complex singlet scalar field (which can also
accommodate a pseudo-Nambu Goldstone dark
matter)~\cite{Cline:2019okt}.\footnote{It should be noted that
in this model the required di-photon decay rate is
reached by adding additional charged particles,
coupling to the 96~GeV scalar.}\addtocounter{footnote}{-1}\addtocounter{Hfootnote}{-1}
\item Composite framework containing a pseudo-Nambu Goldstone-type light
  scalar~\cite{Richard:2020jfd}. 
\item Anomaly-free $U(1)'$ extensions of SM with two complex scalar
  singlets~\cite{AguilarSaavedra:2020wmg}.\footnotemark
\end{itemize}
On the other hand, in the MSSM the CMS excess cannot be
realized~\cite{Bechtle:2016kui}.

%%%%%%%%%%%%%%%%%%%%%%%%%%%%%%%%%%%%%%%%%%%%%%%%%%%%%%%%%%%%%%%%%%%%%%%%%%%%%%
%%%%%%%%%%%%%%%%%%%%%%%%%%%%%%%%%%%%%%%%%%%%%%%%%%%%%%%%%%%%%%%%%%%%%%%%%%%%%%

\section{The N2HDM analysis}

We discussed in~\cite{Biekotter:2019kde}
how a $\sim 96\,$GeV Higgs boson of the Next to minimal 2 Higgs
Doublet Model (N2HDM)~\cite{Chen:2013jvg,Muhlleitner:2016mzt} can be the
origin of both excesses in the type II and type IV scenarios.
The N2HDM extends the CP-conserving 2 Higgs Doublet Model (2HDM) by
a real scalar singlet field. In analogy to the 2HDM, a $Z_2$ symmetry
is imposed to avoid flavor changing neutral currents at the tree level,
which is only softly broken in the Higgs potential. Furthermore, a
second $Z_2$ symmetry, under which the singlet field changes the sign,
constraints the scalar potential. This symmetry is broken spontaneously
as soon as the singlet
field obtains a vacuum expectation value (vev).

In total, the Higgs sector of the N2HDM consists of 3 CP-even Higgs
bosons $h_i$, 1 CP-odd Higgs boson $A$, and 2 charged Higgs bosons
$H^\pm$. In principle, each of the particles $h_i$ can account for the
SM Higgs boson at $125\,$GeV. In our analysis, $h_2$ will be
identified with the SM Higgs boson, while $h_1$ plays the role of the
potential state at $\sim 96\,$GeV. The third CP-even and the CP-odd
states $h_3$ and $A$ were assumed to be heavier than $400\,$GeV to
avoid LHC constraints. The charged Higgs-boson mass was set to be
larger than $650\,$GeV to satisfy constraints from flavor physics
observables.

In the physical basis the 12 independent parameters
of the model are the mixing angles in the CP-even sector
$\alpha_{1,2,3}$, the ratio of the vevs of the Higgs doublets
$\tan\beta=v_2/v_1$, the SM vev $v=\sqrt{v_1^2+v_2^2}$, the vev of the
singlet field $v_S$, the masses of the physical Higgs bosons
$m_{h_{1,2,3}}$, $m_A$ and $M_{H^\pm}$, and the soft $Z_2$ breaking
parameter $m_{12}^2$. 
Using the public code
\texttt{ScannerS}~\cite{Coimbra:2013qq,Muhlleitner:2016mzt}
we performed a scan over the following parameter ranges:
\begin{align}
95 \gev \leq m_{h_1} \leq 98 \gev \; ,
\quad m_{h_2} = 125.09 \gev \; ,
\quad 400 \gev \leq m_{h_3} \leq 1000 \gev \; , \notag \\
400 \gev \leq m_A \leq 1000 \gev \; ,
\quad 650 \gev \leq \MHp \leq 1000 \gev \; , \notag\\
0.5 \leq \tan\beta \leq 4 \; ,
\quad 0 \leq m_{12}^2 \leq 10^6 \gev^2 \; ,
\quad 100 \gev \leq v_S \leq 1500 \gev \; . \label{eq:ranges}
\end{align}

\noindent
The following experimental and theoretical constraints were
taken into account:
\begin{itemize}[noitemsep,topsep=0pt]
\item[-] tree-level perturbativity, boundedness-from-below
	and global-minimum conditions
\item[-] Cross-section limits from collider searches
	using \texttt{HiggsBounds v.5.3.2}~\cite{Bechtle:2008jh,Bechtle:2011sb,Bechtle:2013wla,Bechtle:2015pma}
\item[-] Signal-strength measurements of the SM Higgs boson
	using \texttt{HiggsSignals v.2.2.3}~\cite{Bechtle:2013xfa,Stal:2013hwa,Bechtle:2014ewa}
\item[-] Various flavor physics observables, in particular 
excluding $\MHp < 650\gev$ for all values of 
$\tan\beta$ in the type~II and~IV.
\item[-] Electroweak precision observables in terms
	of the oblique parameters $S$, $T$ and $U$~\cite{Peskin:1990zt,Peskin:1991sw}
\end{itemize}
For more details we refer to~\cite{Biekotter:2019kde}.
The relevant input for \texttt{HiggsBounds} and \texttt{HiggsSignals},
(decay withs, cross sections), were obtained using
the public codes
\texttt{N2HDECAY}~\cite{Muhlleitner:2016mzt,Djouadi:1997yw}
and
\texttt{SusHi}~\cite{Harlander:2012pb,Harlander:2016hcx}.

The results of our parameter scans in the type II and type IV N2HDM,
as given in~\cite{Biekotter:2019kde}, show that both types of the
N2HDM can accommodate the excesses simultaneously, while being in
agreement with all considered constraints described above. A
preference of larger values of $\mu_{\rm CMS}$ in the type II scenario
is visible, which is caused by the suppression of decays into
$\tau$-pairs (see~\cite{Biekotter:2019kde} for details). 

%%%%%%%%%%%%%%%%%%%%%%%%%%%%%%%%%%%%%%%%%%%%%%%%%%%%%%%%%%%%%%%%%%%%%%%%%%%%%%
%%%%%%%%%%%%%%%%%%%%%%%%%%%%%%%%%%%%%%%%%%%%%%%%%%%%%%%%%%%%%%%%%%%%%%%%%%%%%%

\section{Prospects for the LHC Run 3}
\label{sec:run3}

As was shown in \cite{Biekotter:2019kde,Biekotter:2019drs}, the searches
for additional N2HDM Higgs bosons place important constraints on the
allowed parameter space. As discussed above, we employ the code
\texttt{HiggsBounds
  v.5.3.2}~\cite{Bechtle:2008jh,Bechtle:2011sb,Bechtle:2013wla,Bechtle:2015pma}
to apply these searches to the N2HDM parameter space. For each parameter
point \texttt{HiggsBounds} performs the following test. For each Higgs
boson in the model \texttt{HiggsBounds} determines the search channel
with the potentially highest sensitivity by comparing the model
prediction with the {\em expected} limit. Subsequently, only this
channel is used to test the BSM Higgs boson using the {\em observed}
limit on cross section times branching ratio at the 95\%~CL. If the ratio
\begin{align}
  r := \frac{\sigma \times \br(\mbox{predicted})}
            {\sigma \times \br(\mbox{observed limit})}
\end{align}
is found to be larger $\ge 1$, the parameter point is considered
excluded at the 95\%~CL.

Conversely, points with $r < 1$ are not excluded. In general, points
with $r$~below, but close to~1 might be tested in the next round of
experimental data. This is only a rough estimate, since values of
$r$~close to 1 may also be caused by a downward fluctuation of the
background. Nevertheless, we regard points with
\begin{align}
  0.8 \le r < 1.0
\label{rlimit}
\end{align}
as having good chances to be tested in the next round of experimental
analyses.

\begin{figure}[h!]
  \centering
  \includegraphics[width=0.62\textwidth]{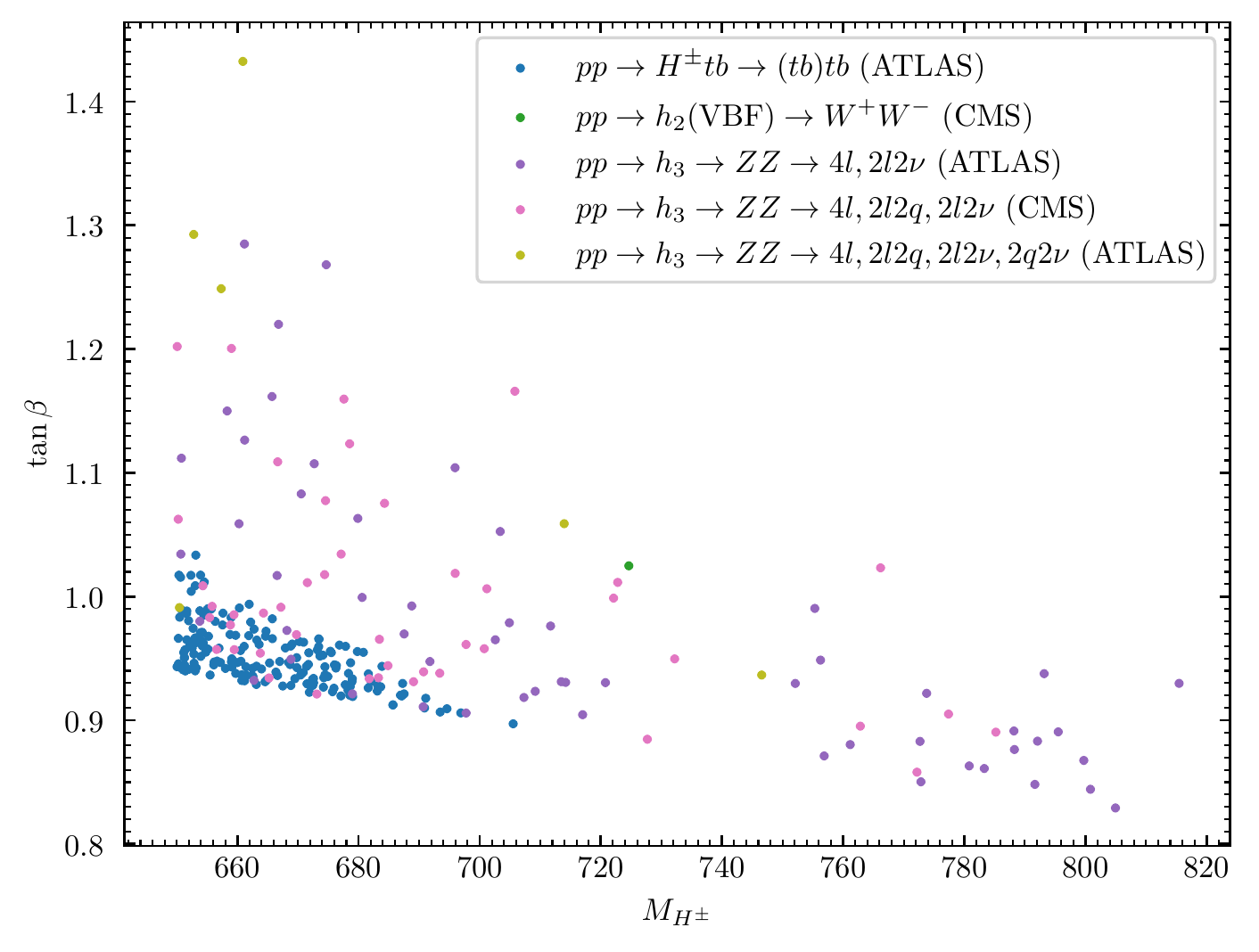}\\
  \includegraphics[width=0.62\textwidth]{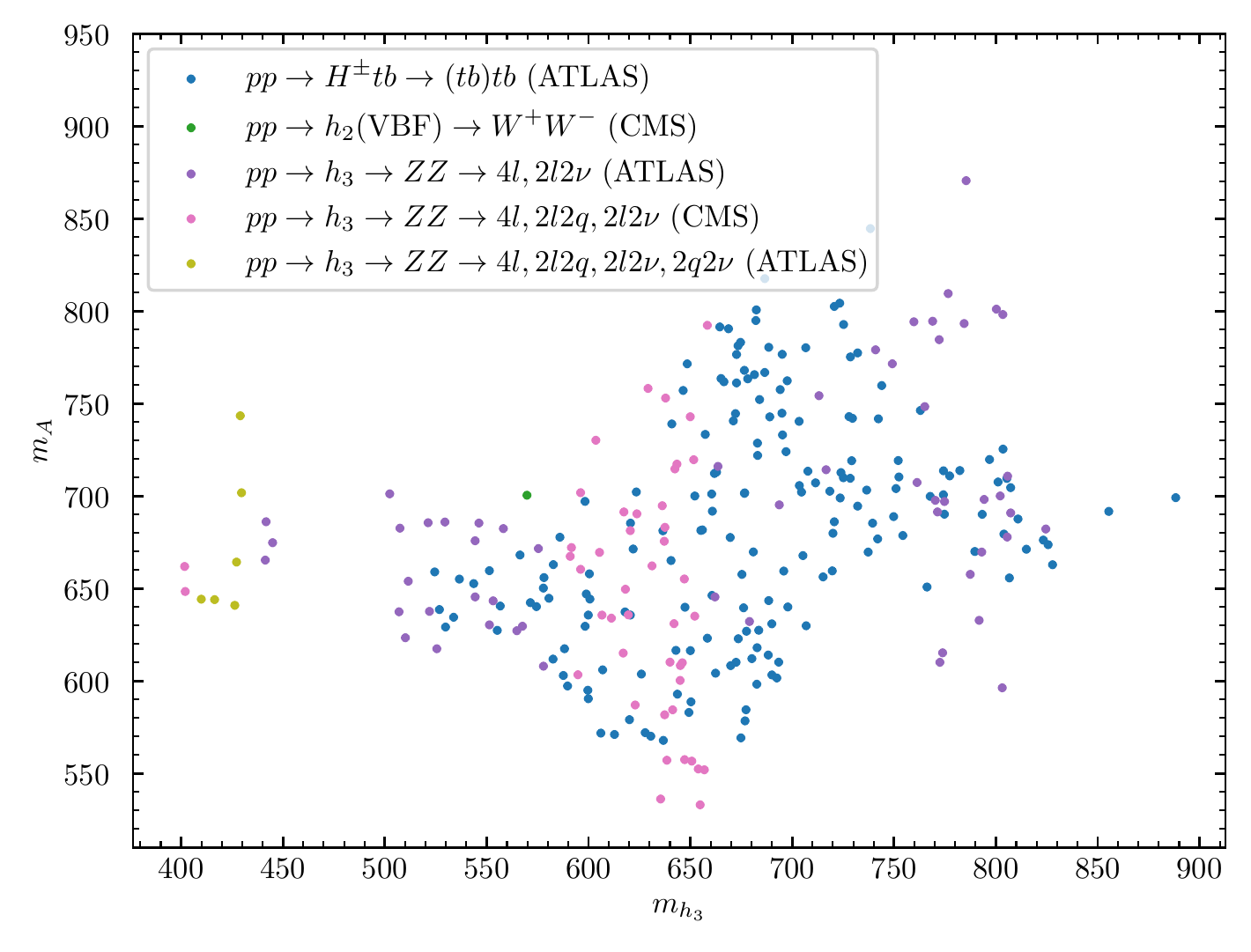}\\
  \includegraphics[width=0.62\textwidth]{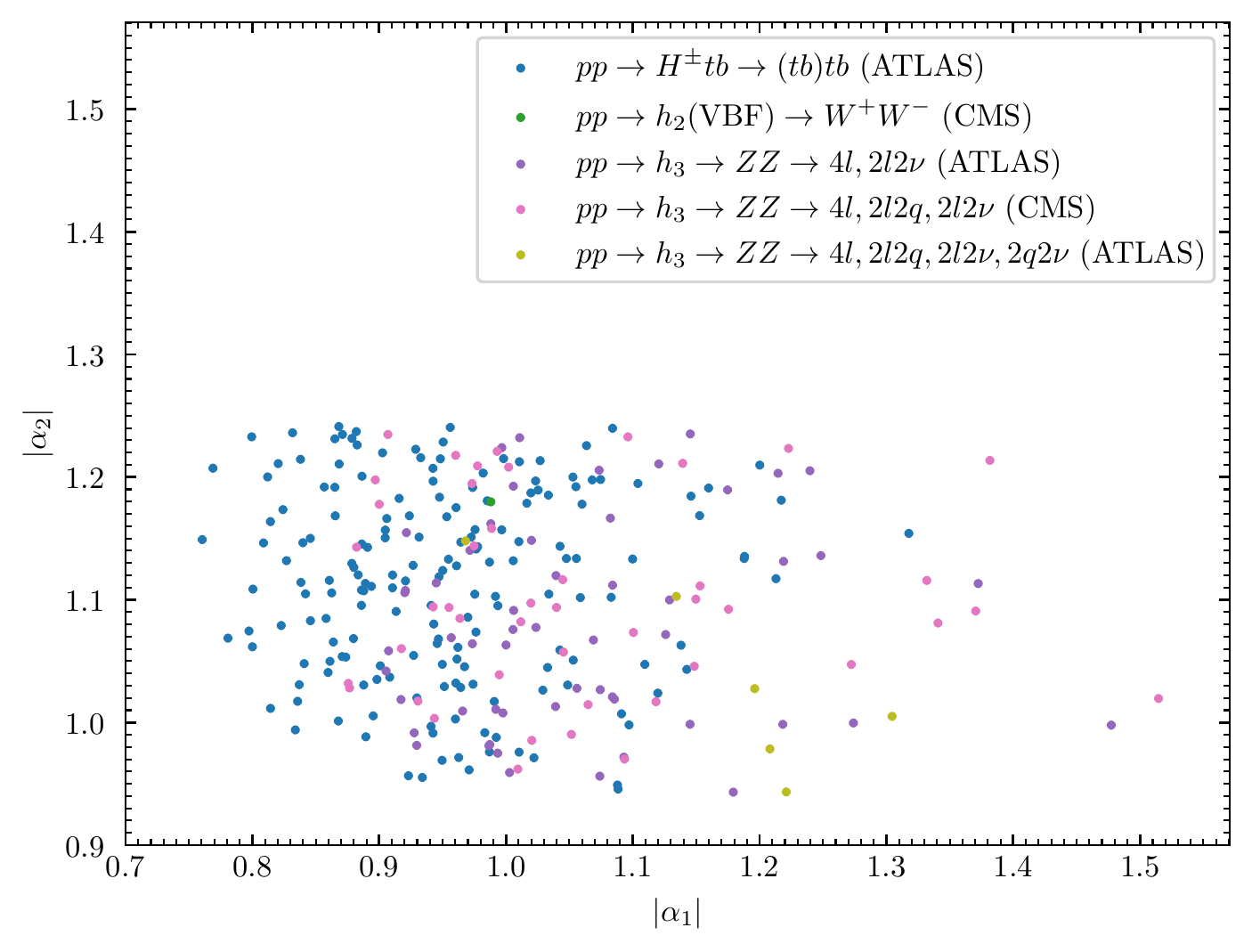}
  \caption{Prospects for the discovery of additional Higgs bosons at the
    LHC Run~3 (see text). The upper, middle and lower plot show the
  planes of $\MHp$-$\tb$, $m_{h_3}$-$m_A$ and $|\alpha_1|$-$|\alpha_2|$,
  respectively.} 
  \label{fig:run3}
\end{figure}

In Fig.~\ref{fig:run3} we show exactly these points
corresponding to Eq.~\ref{rlimit} of our type~II parameter scan, 
focusing on the LHC prospects (i.e.\ disregarding the LEP searches).
The upper, middle and lower plot show the planes of $\MHp$-$\tb$,
$m_{h_3}$-$m_A$ and $|\alpha_1|$-$|\alpha_2|$, respectively.
The color coding denotes the corresponding Higgs boson and its most
senstive search channel. One can identify two main search channels:
\begin{itemize}
\item $pp \to H^\pm tb \to (tb) tb$
\item $pp \to h_3 \to ZZ \to 4$f
\end{itemize}
A third channel, $pp \to h_2 q\bar q \to (W^+W^-) q \bar q$, plays only
a minor role and will not be discussed further.

In the upper plot of Fig.~\ref{fig:run3}, showing the results in the
$\MHp$-$\tb$ plane, one can see that the charged Higgs-boson channel is
relevant at the lowest $\MHp$ values, $\MHp \le 710 \gev$, and at the
lowest $\tb$ values, $\tb \le 1.05$.
This region partially
overlaps with exclusions from flavour physics observables,
which are the origin of the sharp edge of points for even smaller
values of $\tb$.
The parameter space that can be
tested with the heavy neutral Higgs production covers roughly the
triangle up to $\MHp \le 820 \gev$ and $\tb \le 1.3$. However, while
this parameter space does offer interesting possibilities for future
heavy Higgs-boson searches at the LHC, it should be kept in mind that
there are also points in this region with $r < 0.8$ and correspondingly
weaker discovery potential. A detailed analysis of the (HL-)LHC
prospects for the heavy (neutral and charged) Higgs-boson dicovery in
the above denoted channels would be desirable. 

The middle plot of Fig.~\ref{fig:run3} shows the $m_{h_3}$-$m_A$ plane,
following the general pattern based on the analysis of electroweak
precision observables, mainly the $T$~parameter~\cite{Biekotter:2019drs}.
One can observe an approximate separation of the heavy neutral
Higgs-boson decay channels as a function of $m_{h_3}$. The searches for
charged Higgs bosons, on the other hand, are widely distributed in the
allowed parameter space. The bottom plot of Fig.~\ref{fig:run3}, showing
the results in the $|\alpha_1|$-$|\alpha_2|$ plane indicates a rather
uniform distribution of the various Higgs-boson search channel, where
the analysis combining all $h_3 \to ZZ$ decay channels is located at
larger (smaller) values of $|\alpha_1|$ ($|\alpha_2|$).
It should be noted that the upper limit on the $|\alpha_2|$ values
is due to an upper limit on the possible singlet component of the
$96\gev$ scalar in our scans.

%%%%%%%%%%%%%%%%%%%%%%%%%%%%%%%%%%%%%%%%%%%%%%%%%%%%%%%%%%%%%%%%%%%%%%%%%%%%%%
%%%%%%%%%%%%%%%%%%%%%%%%%%%%%%%%%%%%%%%%%%%%%%%%%%%%%%%%%%%%%%%%%%%%%%%%%%%%%%

\section{Conclusion}

The possibility of the N2HDM explaining the CMS and the LEP excess
simultaneously offers interesting prospects to be probed experimentally
in the near future. Here we have focused on the possibilities to
discover additional heavy Higgs bosons in the upcoming LHC runs.
(Prospects for the coupling measurements of the $125 \gev$ Higgs boson
at the HL-LHC and the ILC, as well as the direct production of the
$96 \gev$ Higgs boson at the ILC can be found
in~\cite{Biekotter:2019kde,Biekotter:2020ahz}.)
We find in particular that the searches for the charged Higgs bosons as
well as for the heavy CP-even Higgs boson offer interesting prospects at
the lower allowed values of $\MHp$ and $\tb$. 
A detailed analysis of the (HL-)LHC
prospects for the heavy (neutral and charged) Higgs-boson dicovery in
the above denoted channels would be highly desirable.

%%%%%%%%%%%%%%%%%%%%%%%%%%%%%%%%%%%%%%%%%%%%%%%%%%%%%%%%%%%%%%%%%%%%%%%%%%%%%%
%%%%%%%%%%%%%%%%%%%%%%%%%%%%%%%%%%%%%%%%%%%%%%%%%%%%%%%%%%%%%%%%%%%%%%%%%%%%%%

\section*{Acknowledgements}

The work  was supported in part by the MEINCOP (Spain) under 
contract FPA2016-78022-P and in part by the AEI
through the grant IFT Centro de Excelencia Severo Ochoa SEV-2016-0597. 
The work of S.H.\ was 
supported in part by the Spanish Agencia Estatal de
Investigaci\'on (AEI), in part by
the EU Fondo Europeo de Desarrollo Regional (FEDER) through the project
FPA2016-78645-P, in part by the ``Spanish Red Consolider MultiDark''
FPA2017-90566-REDC.
T.B.\ is supported by the Deutsche Forschungsgemeinschaft under
Germany's Excellence Strategy EXC2121 ``Quantum Universe'' - 390833306.
% The work of T.B. was
% funded by Fundaci\'on La Caixa under `La Caixa-Severo Ochoa' international
% predoctoral grant.

\section*{References}
\bibliography{hongkong}

\end{document}